\documentclass[journal]{IEEEtran}

\usepackage[T1]{fontenc}
\usepackage[utf8]{inputenc}
\usepackage{amsmath,amssymb,amsfonts}
\usepackage{graphicx}
\usepackage{textcomp}
\usepackage{xcolor}
\usepackage{hyperref}
\usepackage{cite}

\begin{document}

\title{Detecting Outliers of Pursuit Eye Movements: A Preliminary Analysis of Autism Spectrum Disorder}

\author{Emiko~Shishido$^{a,*}$,
        Seiko~Miyata$^{b}$,
        Tetsuya~Yamamoto$^{a}$,
        Masaki~Fukunaga$^{a}$,
        Ryota~Hashimoto$^{c}$,
        Kenichiro~Miura$^{d}$,
        and~Norio~Ozaki$^{e}$

\thanks{$^{a}$National Institute for Physiological Sciences, Okazaki, Japan.}
\thanks{$^{b}$Department of Psychiatry, Nagoya University, Graduate School of Medicine, Nagoya, Aichi, Japan.}
\thanks{$^{c}$Department of Pathology of Mental Diseases, National Institute of Mental Health, National Center of Neurology and Psychiatry, Kodaira, Tokyo, Japan.}
\thanks{$^{d}$Institute for the Advanced Study of Human Biology, Kyoto University, Kyoto, Japan.}
\thanks{$^{e}$Pathophysiology of Mental Disorders, Nagoya University Graduate School of Medicine, Nagoya, Aichi, Japan.}
\thanks{$^{*}$Corresponding author: Emiko Shishido, Ph.D., National Institute for Physiological Sciences, Okazaki, Japan. Email: emikosh[at mark]nips.ac.jp}
\thanks{This research was supported by AMED grant JP23ak0101215 and JSPS KAKENHI Grant Number 24K10723.}
}

\markboth{Outliers of Pursuit Eye Movement}{}

\maketitle

\begin{abstract}
\textbf{Background:}
Autism spectrum disorder (ASD) is characterized by significant clinical and biological heterogeneity. Conventional group-mean analyses of eye movements often mask individual atypicalities, potentially overlooking critical pathological signatures. This study aimed to identify idiosyncratic oculomotor patterns in ASD using an ``outlier analysis'' of smooth pursuit eye movement (SPEM).

\textbf{Methods:}
We recorded SPEM during a slow Lissajous pursuit task in 18 adults with ASD and 39 typically developed (TD) individuals. To quantify individual deviations, we derived an ``outlier score'' based on the Mahalanobis distance. This score was calculated from a feature vector, optimized via Principal Component Analysis (PCA), comprising the temporal lag ($\Delta t$) and the spatial deviation ($\Delta s$). An outlier was statistically defined as a score exceeding $\sqrt{10}$ (approximately 3.16\,$\sigma$) relative to the TD normative distribution.

\textbf{Results:}
While the TD group exhibited a low outlier rate of 5.1\% (2/39), the ASD group demonstrated a significantly higher prevalence of 38.9\% (7/18) (binomial $P = 0.0034$). Furthermore, the mean outlier score was significantly elevated in the ASD group ($3.00 \pm 2.62$) compared to the TD group ($1.52 \pm 0.80$; $P = 0.002$). Notably, these extreme deviations were captured even when conventional mean-based comparisons showed limited sensitivity.

\textbf{Conclusions:}
Our outlier analysis successfully visualized the high degree of idiosyncratic atypicality in ASD oculomotor control. By shifting the focus from group averages to individual deviations, this approach provides a sensitive metric for capturing the inherent heterogeneity of ASD, offering a potential baseline for identifying clinical subtypes.
\end{abstract}

\begin{IEEEkeywords}
autism spectrum disorder, eye tracking, smooth pursuit, outlier analysis
\end{IEEEkeywords}

\section{Introduction}

The pathophysiology of psychiatric disorders remains complex and poorly understood. Establishing objective, quantifiable biomarkers is crucial for improving diagnostic accuracy and developing personalized interventions. Among various candidates, eye-tracking has emerged as a promising non-invasive tool, offering insights into neurocognitive functions with relatively simple experimental setups compared to neuroimaging techniques like MRI. Recent studies have demonstrated that integrating multiple oculomotor indices can significantly enhance diagnostic performance. For instance, a combination of eye movement parameters has been shown to distinguish individuals with schizophrenia (SZ) from healthy controls with high precision \cite{Miura2014, Morita2017, Zhu2024}.

While ASD also exhibits distinct oculomotor profiles---such as reduced smooth pursuit eye movement (SPEM) accuracy and atypical saccade patterns---these findings are often less pronounced than those in SZ and show significant variability across studies \cite{Caldani2023, Miyata2026, Shiino2020}. Across candidate studies, ASD cohorts show reduced pursuit maintenance or gain, greater saccade amplitude variability, and poorer fixation stability, with substantial developmental and methodological sensitivity.

The heterogeneity of ASD has long been recognized as a significant barrier to establishing consistent biomarkers. Traditional case-control studies focusing on group differences often mask the diverse neurobiological signatures present in only a subset of individuals. To overcome this, Campbell et al.\ \cite{Campbell2013} introduced a pathway-based outlier analysis using Mahalanobis distance to characterize idiosyncratic genomic structures in ASD, demonstrating that significant biological signals can be identified by focusing on individuals who deviate from the normative distribution.

In the present study, we applied outlier analysis to SPEM data from an adult ASD cohort \cite{Miyata2026}. The ``outlier score'' was defined as the Mahalanobis distance calculated from the principal components of three spatial and temporal deviation variables. Unlike traditional group-comparison approaches that focus on mean differences, this method identifies individuals whose oculomotor profiles deviate significantly from the normative distribution of the TD group. This preliminary analysis seeks to bridge the gap between group-level findings and individualized clinical assessment, potentially providing a new metric for the objective quantification of ASD-related neurocognitive traits.

\section{Materials and Methods}

\subsection{Participants}

This study included 18 individuals with ASD and 39 age-matched TD individuals, recruited from visitors and staff of Nagoya University Hospital from 2015 to 2019. Inclusion and exclusion criteria are described in Miyata et al.\ \cite{Miyata2026}. The study was performed in accordance with the tenets of the Declaration of Helsinki of the World Medical Association and was approved by the Research Ethical Committee of Nagoya University (No.\ 2010-0930). All participants provided written informed consent after receiving a comprehensive explanation of the study procedures.

\subsection{Outlier Analysis of Oculomotor Data}

\begin{figure} [ht!]
\centering
\includegraphics[width=\columnwidth]{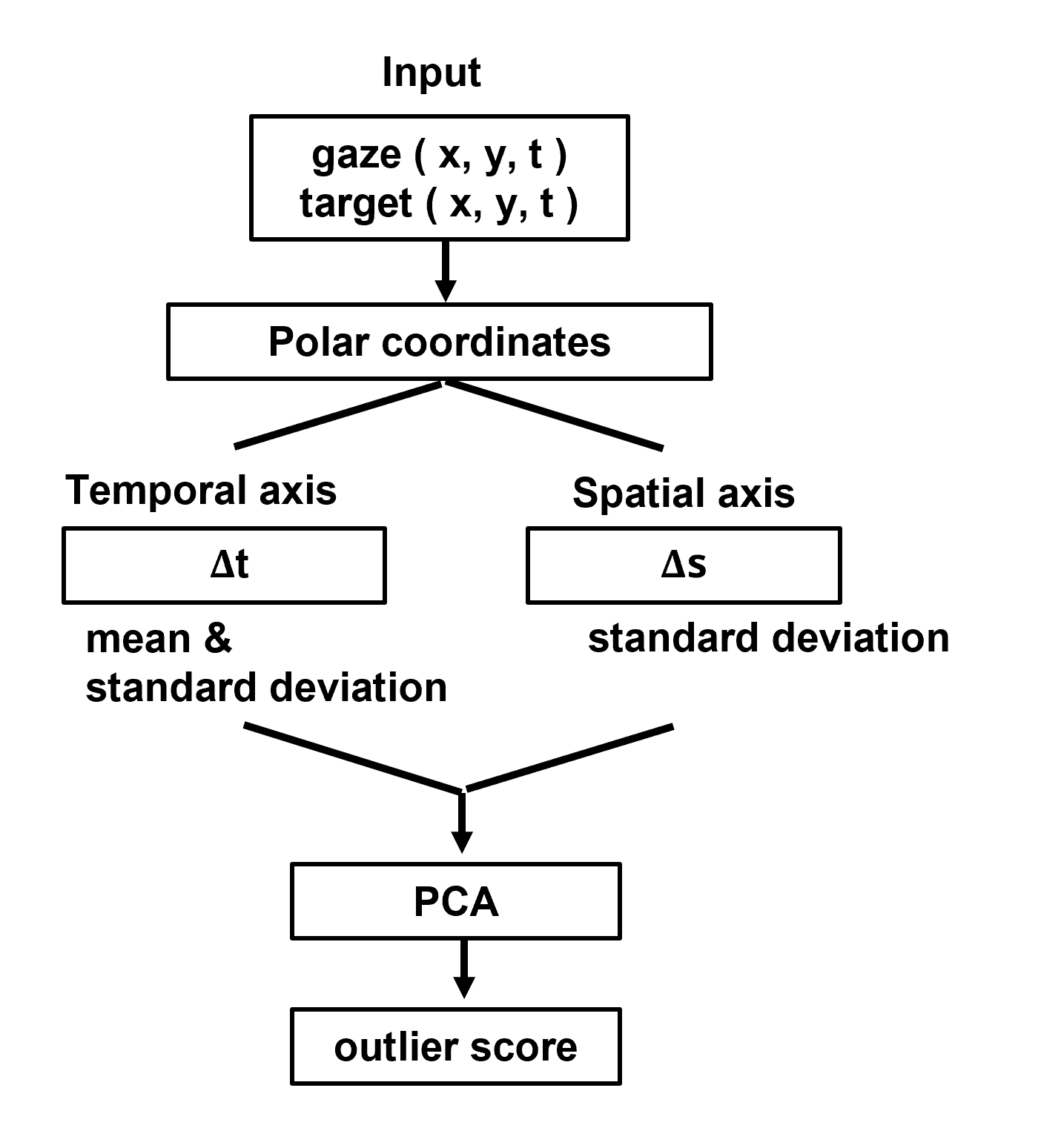}
\caption{Analytical pipeline for the calculation of the outlier score.
This flowchart illustrates the step-by-step process used to quantify individual atypicality in smooth pursuit eye movement (SPEM).
\textbf{Input Data}: High-resolution gaze and target trajectories $(x, y, t)$ were recorded.
\textbf{Polar Coordinate Transformation}: Raw coordinates were converted into a polar coordinate system relative to the display center.
\textbf{Temporal and Spatial Axis Analysis}:
  Temporal axis ($\Delta t$): calculated as the time difference for the gaze to reach the same radial position as the target; features include the mean and standard deviation of $\Delta t$.
  Spatial axis ($\Delta s$): calculated as the difference in radial distance, normalized by the instantaneous target radius; only the standard deviation was utilized.
\textbf{Outlier Score Calculation}: The three parameters were integrated via PCA and Mahalanobis distance to derive the final ``outlier score,'' representing the degree of departure from the TD normative pattern.}
\label{fig:pipeline}
\end{figure}

Data processing was performed using MATLAB (R2023b). The outlier score was defined as the Mahalanobis distance derived from a principal component analysis (PCA) of three initial features: the mean and standard deviation of temporal deviation ($\Delta t$, Fig.~\ref{fig:pipeline}) and the standard deviation of the normalized spatial deviation ($\Delta s$, Fig.~\ref{fig:pipeline}). By reducing dimensionality based on variance contribution, the final feature vector effectively captured individual idiosyncratic patterns. The normative reference was established using TD group data. A higher outlier score indicates a greater degree of deviation from the typical oculomotor pattern. Detailed mathematical procedures are provided in the \href{https://github.com/emiko-sh/outlier_ASD_TD_20260323/tree/main}{Supplementary Methods}.

\subsection{Age and Outlier Scores}

Statistical analysis was performed using MATLAB (R2023b). To account for the potential influence of aging on oculomotor performance, we examined the correlation between age and outlier scores in the TD group using Pearson's correlation analysis. Group differences in outlier scores between the TD and ASD groups were compared using non-paired two-tailed $t$-tests. For all analyses, the significance level was set at $p < 0.05$.

\subsection{Probability of Outliers}

To evaluate whether the number of outliers in the ASD group was statistically significant, we employed interval estimation of binomial parameters. We defined the threshold for an ``outlier'' as a score exceeding $\sqrt{10}$ (approximately 3.16 SD), representing extreme deviation from the normative TD mean. Using the 95\% confidence interval of the outlier rate observed in the TD group, we calculated the probability ($p$-value) of observing the actual number of ASD outliers under the null hypothesis. This analysis allowed us to determine whether the overrepresentation of outliers in the ASD group was a chance occurrence or a reflection of the disorder's inherent neurocognitive atypicality.

\section{Results}

\subsection{Evaluation of Individual Deviations via Outlier Score}

\begin{figure}[!t]
\centering
\includegraphics[width=\columnwidth]{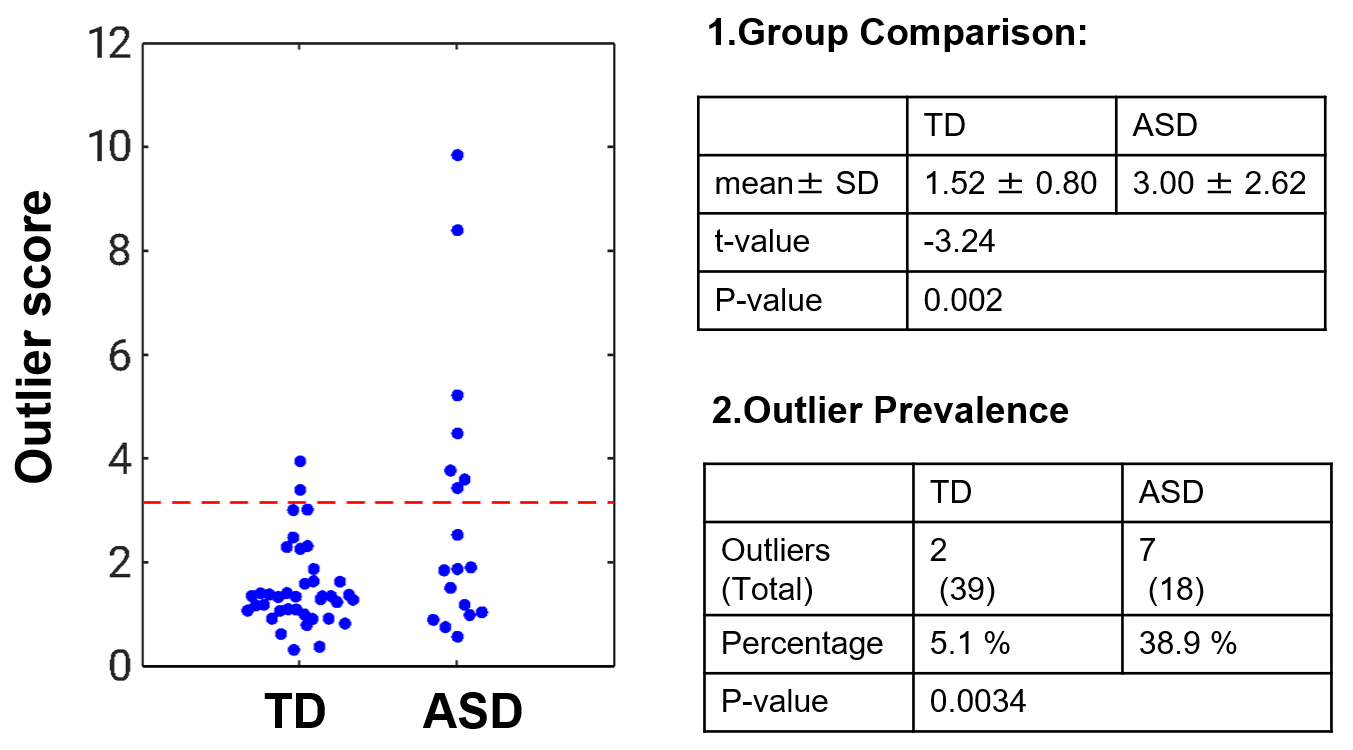}
\caption{Outlier scores of typically developed (TD) individuals and individuals with autism spectrum disorder (ASD).
\textbf{(Left)} Scatter plot of Outlier Scores: each dot represents the outlier score for an individual participant, calculated as the Mahalanobis distance derived from PCA of temporal and spatial deviation features. The dashed red line indicates the statistical threshold for outliers ($\sqrt{10} \approx 3.16$).
\textbf{(Right)} Statistical summary: the ASD group exhibited significantly higher mean outlier scores ($3.00 \pm 2.62$) compared to the TD group ($1.52 \pm 0.80$; $P = 0.002$). The TD group showed a low outlier rate of 5.1\% (2/39), whereas the ASD group demonstrated a markedly higher prevalence of 38.9\% (7/18). A binomial parameter analysis confirmed that the proportion of outliers in the ASD group was significantly greater than that in the TD group ($P = 0.0034$).}
\label{fig:outlier_scores}
\end{figure}

To more effectively characterize the idiosyncratic oculomotor patterns in ASD, we applied an outlier analysis based on point-by-point subtraction of temporal ($\Delta t$) and spatial ($\Delta s$) deviations in a polar coordinate system. The comparison of composite Outlier Scores revealed a striking difference between the groups (Fig.~\ref{fig:outlier_scores}). The ASD group exhibited significantly higher Outlier Scores compared to the TD group ($P = 0.002$). While the TD group's scores were narrowly clustered, indicating a consistent ``typical'' pursuit pattern, the ASD group's scores showed a broad distribution. Several ASD individuals exhibited scores multiple standard deviations away from the TD mean, reflecting a high degree of individual ``atypicality'' not captured by traditional averaging methods.

\subsection{Statistical Significance of Outlier Frequency}

To determine whether the prevalence of outliers in the ASD group significantly exceeded what would be expected by chance, we performed an analysis using interval estimation of binomial parameters. Based on the normative distribution of the TD group, we established a stringent threshold at $\sqrt{10}$ (approximately 3.16\,$\sigma$). While only 5.1\% (2/39) of the TD group exceeded this threshold, 38.9\% (7/18) of the ASD group were identified as outliers, confirming a distinct distributional shift in the clinical population.

We estimated the 95\% confidence interval for the outlier rate within the TD population. A subsequent binomial test revealed that the number of individuals in the ASD group exceeding this threshold was significantly higher than the estimated rate for the TD group ($P = 0.0034$). This result indicates that the atypical oculomotor patterns observed in the ASD cohort represent a statistically significant overrepresentation of extreme physiological deviations, not merely random variations.

\subsection{Relationship with Age}

Correlation analyses were conducted to evaluate the influence of aging on oculomotor stability. In the TD group, there was no significant correlation between age and the Outlier Score ($R = -0.25$, $P = 0.99$), suggesting that the Outlier Score is relatively stable across the age range studied. Similarly, the ASD group did not show a significant age-related trend in Outlier Scores ($R = -0.16$, $P = 0.86$). These results indicate that the elevated Outlier Scores in the ASD group are primarily reflective of the disorder's neurocognitive characteristics rather than age-dependent decline.

\section{Discussion}

In the present study, we applied a novel outlier analysis based on point-by-point spatial and temporal deviations to characterize the idiosyncratic SPEM patterns in adults with ASD. Our findings revealed that while traditional group-mean comparisons provide a foundational understanding of oculomotor characteristics \cite{Miyata2026}, the outlier score offers a more sensitive and individualized metric for capturing the inherent heterogeneity of ASD.

\subsection{Beyond Group Means: The Significance of Outlier Analysis}

As previously reported \cite{Miyata2026}, the COCORO standardized protocol has proven effective in identifying group-level differences between ASD and TD. However, as noted in recent literature, the effect sizes of oculomotor impairments in ASD are often smaller and more variable than those observed in schizophrenia \cite{Thomas2021}. This variability can lead to inconsistent results across studies when relying solely on mean-based statistics.

By calculating spatial ($\Delta s$) and temporal ($\Delta t$) deviations in a polar coordinate system for Lissajous trajectories, we successfully quantified ``atypicality'' that might otherwise be masked by averaging. The significantly higher and more dispersed outlier scores in the ASD group suggest that the disorder is characterized not by a single ``ASD-type'' eye movement, but by a diverse range of deviations from the normative typical distribution.

Our findings that the outlier score captured significant differences where traditional group-mean comparisons failed are consistent with previous genomic research. Campbell et al.\ \cite{Campbell2013} demonstrated that while group-level differential expression analysis (e.g., GSEA) often yields non-significant results in ASD, an outlier-based approach can reveal distinct subgroups with specific biological perturbations. By applying similar logic to oculomotor dynamics---specifically by quantifying point-by-point temporal and spatial deviations---we were able to identify `atypical' pursuit patterns that characterize the idiosyncratic nature of ASD. This underscores the importance of shifting the analytical focus from `group averages' to `individual deviations' in the study of heterogeneous neurodevelopmental disorders.

\subsection{Spatial vs.\ Temporal Impairments in SPEM}

A key technical advantage of our methodology is the high-resolution subtraction of gaze position from the target trajectory. The integration of spatial precision and temporal synchronization reflects the complex neural processing required for smooth pursuit, involving the prefrontal cortex and cerebellum. Our results suggest that some ASD individuals exhibit a primary ``lag'' (temporal deviation), while others show ``spatial drift'' (spatial deviation). Such fine-grained differentiation is crucial for understanding the diverse neurobiological backgrounds of ASD, which is often difficult to achieve with conventional indices like gain or SNR alone.

\subsection{Clinical Implications and Future Directions}

The ability to objectively quantify an individual's deviation from the typical population has significant clinical potential. The outlier score could serve as a supplementary diagnostic tool or a metric for stratifying ASD subgroups based on neurocognitive profiles. Furthermore, the lack of significant correlation between outlier scores and age in both groups indicates that these oculomotor signatures are stable traits, potentially serving as reliable biomarkers.

\subsection{Limitations}

Despite the increased sensitivity of the outlier analysis, the sample size remains relatively small. Further validation in larger, independent cohorts is necessary to confirm the diagnostic utility of the outlier score. Future research should also explore the correlation between these oculomotor outliers and specific clinical dimensions, such as sensory hypersensitivity or social communication scores.

\section*{Acknowledgments}

We appreciate Nanayo Ogawa for recruiting participants and providing WAIS scores, and Chikako Kawai for technical support. We thank Dr.\ Norihiro Sadato for helpful discussions. This research was supported by AMED grant JP23ak0101215 and JSPS KAKENHI Grant Number 24K10723.

\section*{Author Contributions}

\textbf{Emiko Shishido}: Writing -- Original Draft, Investigation, Methodology, Formal Analysis, Writing -- Review \& Editing.
\textbf{Seiko Miyata}: Writing -- Review.
\textbf{Tetsuya Yamamoto}: Methodology.
\textbf{Masaki Fukunaga}: Methodology, Writing -- Review.
\textbf{Ryota Hashimoto}: Writing -- Review.
\textbf{Kenichiro Miura}: Methodology, Writing -- Review.
\textbf{Norio Ozaki}: Funding Acquisition, Writing -- Review, Supervision.

\section*{Conflicts of Interest}

The authors have no relevant financial interests to disclose.
\section*{Supplementary Materials}
\begin{itemize}
  \item \url{score_data.xlsx} :\\
  List of outlier score.
  \item \url{Supplementary Table.pdf} :\\
  Traditional eye movement score (``16 EM score''). 
\end{itemize}

\section*{Supplementary Methods}
Detailed supplementary methods and analysis scripts are available at the following GitHub repository: \\
\url{https://github.com/emiko-sh/outlier_ASD_TD_20260323/tree/main}

\section*{Supplementary Movies}
\begin{itemize}
    \item Media1.mp4 (TD group): \\
    A representative example of a typical smooth pursuit pattern from the TD group (Outlier score: 1.64).
    \item Media2.mp4 (ASD group): \\
    An ``outlier'' pattern from the ASD group, demonstrating significant spatial drift and temporal lag (Outlier score: 5.22).
\end{itemize}

\section*{Citation}
Shishido, E., et al. (2026). Detecting outliers of pursuit eye movements: a preliminary analysis of autism spectrum disorder. arXiv: https://arxiv.org/pdf/2603.22705\\



\end{document}